\documentclass[twocolumn,showpacs,prl]{revtex4}
\usepackage{amsmath}
\usepackage{epsf}
\usepackage{graphicx}
\usepackage{dcolumn}
\usepackage{bm}

\begin{document}

\title{Interacting fermions in two dimensions: beyond the perturbation theory}
\author{Suhas Gangadharaiah$^{1}$, Dmitrii L. Maslov$^{1*,2}$, Andrey V.
Chubukov$^{3}$, and Leonid I. Glazman$^{4}$}
\date{\today}

\begin{abstract}
We consider a system of 2D fermions with short-range interaction. A
straightforward perturbation theory is shown to be ill-defined even for an
infinitesimally weak interaction, as the perturbative series for the
self-energy diverges near the mass shell. We show that the divergences
result from the interaction of fermions with the zero-sound collective mode.
By re-summing the most divergent diagrams, we obtain a closed form of the
self-energy near the mass shell. The spectral function exhibits a threshold
feature at the onset of the emission of the zero-sound waves. We also show
that the interaction with the zero sound does not affect a non-analytic, $%
T^{2}$-part of the specific heat.
\end{abstract}

\affiliation{ $^{1}$Department of
Physics, University of Florida, P. O. Box 118440, Gainesville, FL
32611-8440 \\
$^{2}$Abdus Salam International Center for Theoretical Physics, 11
Strada Costiera, 34014
Trieste, Italy\\
$^3$Department of Physics, University of Maryland,
College Park, MD 20742-4111\\
$^4$Theoretical Physics Institute, University of Minnesota,
Minneapolis, MN 55455}

\maketitle

 The Landau Fermi-liquid (FL) theory states that the
low-energy properties of an interacting fermion system are similar
to those of an ideal Fermi gas \cite{agd}. A necessary (but not
sufficient) condition for the validity of this theory is that
various characteristic properties (thermodynamic parameters,
quasi-particle lifetime, etc.) can be expressed via regular
perturbative expansions in the interaction. Back in the 50s, it
was shown to be the case for a model of 3D fermions with both
short- and long-range (Coulomb) repulsion. Soon thereafter, it was
reazlied that the perturbation
theory is singular in 1D and, as a result, the FL is destroyed. The case of $%
D=2$ had been the subject of an active and fairly recent discussion, with a
number of proposals, most notably by Anderson \cite{anderson}, for the
breakdown of the FL in 2D. Although the prevailing opinion currently is that
the FL is stable in 2D for sufficiently short-range (including Coulomb)
interactions, a full description of the FL in 2D is still lacking.

One of the problems in describing a 2D interacting system is that, in a
contrast to the 3D case, a naive perturbative expansion in the interaction
is singular. Namely, for a linearized single-particle spectrum,
the imaginary part of the self-energy
diverges on the mass shell.
The divergence is logarithmic to second order in the interaction \cite
{castellani,fukuyama,metzner,chm}
 but, as we will show in this paper, it is amplified to a power-law, starting
from the third order. This mass-shell singularity in 2D is a weaker form of the ``infrared
catastrophe'' in 1D.
There, an on-shell fermion can emit an infinite number of soft
bosons--quanta of charge- and spin-density fluctuations \cite{1D}, which
gives rise to a power-law divergence of the self-energy on the mass-shell.
In 2D, this mechanism is weakened but not
completely eliminated (the ``memory'' about the 1D infrared catastrophe is
erased completely only for $D>2$). At a first glance, the breakdown of the
perturbation theory confirms the conjecture that the FL is destroyed in 2D
\cite{anderson}. However, all it really means is that in order to obtain
physically meaningful results the perturbation theory must be re-summed even
for an \emph{arbitrarily weak interaction} \cite{castellani,metzner}.

In this Letter, we report the results of an asymptotically exact
re-summation of the divergent perturbation theory for $D=2$. In
addition to obtaining a closed and divergence-free form of the
self-energy to all orders in the interaction, this procedure also
allows one to identify the nature of the singularities in a
perturbation theory as originating from the interaction between
the fermions and the zero-sound (ZS) collective mode. At any
finite order of the perturbation theory, the collective mode
coincides with the upper edge of the particle-hole (PH) continuum.
In 2D, this degeneracy is strong enough to generate the
divergences in the self-energy. Once perturbations are summed up
to all orders, the zero-sound mode splits off from the continuum,
and the power-law divergences disappear. The remaining
log-divergences are eliminated by restoring the finite curvature
of the spectrum near the Fermi surface \cite{metzner,chm}.
 Thus the FL survives. However, the resulting
self-energy contains several non-perturbative features that signal
a deviation from the standard FL behavior. First, the imaginary
part of a non-perturbative, ZS contribution to the self-energy,
$\Sigma _{\mathrm{ZS}}\left( \omega ,k\right)$, is a non-monotonic
function of the ``distance'' to the mass-shell, $\Delta \equiv
\omega -\epsilon _{k}$, and exhibits an anomaly at the threshold
for emission of ZS waves. This anomaly gives rise to a
non-Lorentzian shape of the spectral function. Second, as the
theory is regularized by the difference in the ZS and Fermi
velocities, which is small for a weak interaction, $\Sigma
_{\mathrm{ZS}}$ is strongly enhanced near the mass
shell. On the mass shell, $\mathrm{Re}\Sigma _{{\rm ZS}}$ is of the same --
$U^{2}\omega |\omega |$--order, as the perturbative contribution.
 It has been argued that the non-analytic,
perturbative self-energy Re$\Sigma \propto U^{2}\omega |\omega|$
 gives rise to a non-analytic, $T^{2}$-term in
the specific heat $C(T)$~\cite{bedell,chm,dassarma}.
Therefore, it becomes an issue whether the
non-perturbative part of the self-energy also contributes to the $T^{2}$
-term in $C(T)$. We show that this is not the case, as the
thermodynamic potential is not affected by the mass-shell singularities.

In what follows, we present the results and outline the
main steps. Detailed calculations can be found in the extended
version of this communication \cite{long}.
\begin{figure}[tbp]
\begin{center}
\epsfxsize=0.8 \columnwidth \epsffile{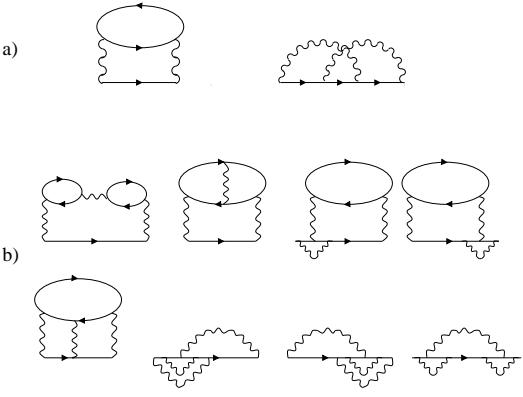}
\end{center}
\caption{Maximally divergent self-energy diagrams to second (a)
and third (b) orders in the interaction. } \label{fig:sigma}
\end{figure}

\begin{figure}[tbp]
\begin{center}
\epsfxsize=1.0 \columnwidth \epsffile{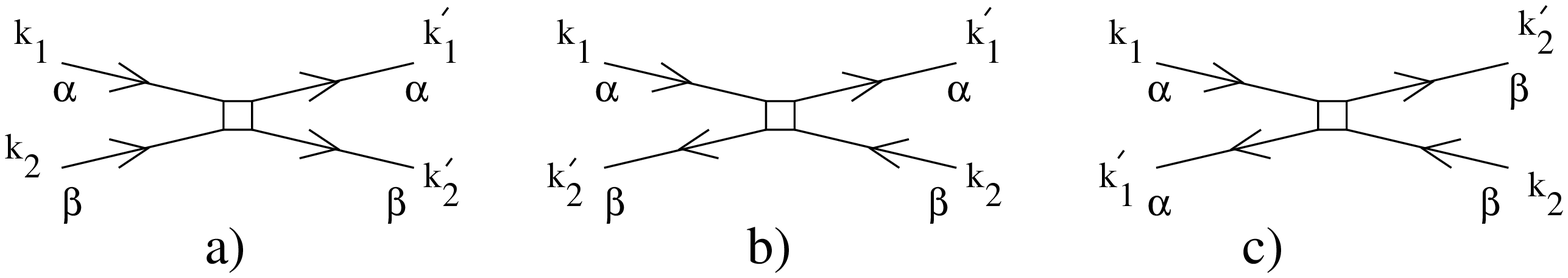}
\end{center}
\caption{Forward(a) and backscattering (b+c) scattering
processes.} \label{fig:proc}
\end{figure}

To understand the origin of the mass-shell singularity in
arbitrary dimension $D$, consider the self-energy to  second
order in the interaction (first two diagrams in
Fig.\ref{fig:sigma}), and focus on small-angle scattering of
fermions with almost parallel momenta (forward scattering), as
shown in Fig.~\ref{fig:proc}a.

The imaginary part of the retarded self-energy on the mass shell
is given by (for $\omega>0$)
\begin{eqnarray}
\text{Im}\Sigma _{\mathrm{F}}(\omega =\epsilon _{k}) &\propto
&u_{0}^{2}\int_{-\omega}^{0 }d\Omega \int dQQ^{D-1}\int dO_{D}  \notag \\
&&\times \delta \left( \Omega +v_{F}Q\cos \vartheta \right) \text{Im}\Pi
(\Omega ,Q),  \label{a}
\end{eqnarray}
where $u_{0}\equiv U\left( 0\right) m/2\pi $ is the dimensionless coupling
constant, $\Pi (\Omega ,Q)$ is the polarization bubble, and  $dO_{1}=\delta
(\vartheta )d\vartheta $, $dO_{2}=d\vartheta $, $dO_{3}=\sin \vartheta
d\vartheta $. $\mathrm{Im}\Pi (\Omega ,Q)$ is the probability
amplitude to generate a PH pair of frequency $\Omega $ and momentum $Q$. For
$D=1$, the PH continuum consists of a single line $\Omega=v_{F}Q$ and $%
\mathrm{Im}\Pi (\Omega ,Q)\propto \delta (\Omega -v_{F}Q)$. A
product of two $\delta $-functions in (\ref{a}) yields a
non-integrable singularity, which is the origin of an infrared
catastrophe \cite{1D}. For $D>1$, the PH continuum is a broad band
specified by $|\Omega | \leq v_{F}Q$, and the fermion spectral
function is softened by  angle-averaging. For $D=3$, the resulting
self-energy is finite on the mass shell. However, for $D=2$ the
angle-averaged fermion spectral function still has a square- root
singularity: $\int d\theta \delta (\Omega -v_{F}Q\cos \vartheta
)\propto
\left( v_{F}Q-|\Omega | \right) ^{-1/2}$ for $v_{F}Q>|\Omega |$, whereas $\mathrm{%
Im}\Pi (\Omega ,Q)\propto \Omega /\left( v_{F}Q-|\Omega |\right) ^{1/2}$ has
another square-root singularity at the continuum boundary. Merging of the
two square-root singularities results in a logarithmic divergence of $%
\mathrm{Im}\Sigma $ on the mass shell. Near the mass shell ($|\Delta |\ll
|\omega |$), $\mathrm{Im}\Sigma \propto \ln |\Delta |$.

Higher-order diagrams contain higher powers of divergent PH
bubbles; these bubbles are either explicit, as in the ring
diagrams of Fig.~\ref{fig:sigma}, or generated by integrating out
the fermionic energy and momenta. One can verify that the
square-root singularities accumulate, and the
higher-order diagrams for the self-energy diverge as powers of $\Delta ^{-1}$%
. At the n-th order, $\text{Im}\Sigma _{F}\propto u_{0}^{n}\omega
^{2}\left( \omega /\Delta \right) ^{n/2-1}\theta (\omega /\Delta
)$, where $\theta (x)$ is the step-function. By Kramers-Kronig
relation, these singularities generate similar power-law
divergences in $\text{Re}\Sigma _{F}$. Hence, to find the
self-energy in $D=2$ one has to re-sum the perturbative series
even for an infinitesimally small $U.$

\begin{figure}[tbp]
\begin{center}
\epsfxsize=0.6 \columnwidth
\epsffile{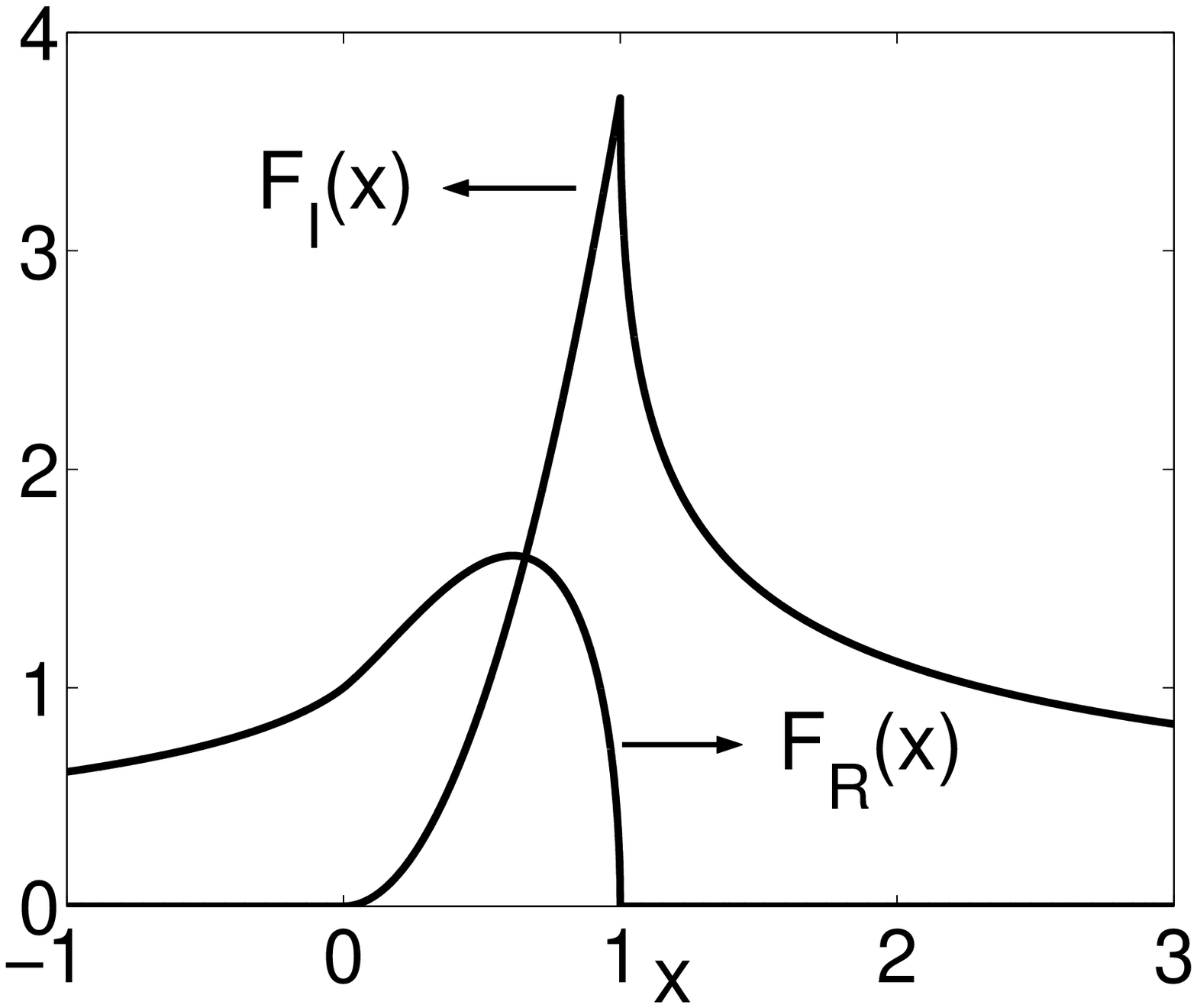}
\end{center}
\caption{Scaling functions $F_{I}\left( x\right) $ and
$F_{R}\left( x\right) $} \label{fig:scaling}
\end{figure}

This summation can be carried out explicitly for small $U$ by
collecting diagrams with the maximum number of polarization
bubbles at each order (to third order, such diagrams are shown in
Fig.~\ref{fig:sigma}). The result is
\begin{eqnarray}
&&\Sigma _{F}\left( p\right) =\frac{1}{2}~\int_{q}~G\left( p-q\right) \Bigg [%
4U(0)-2U^{2}(0)\Pi (q)  \notag \\
+ &&\frac{U(0)}{1-U(0)\Pi \left( q\right) }-\frac{3U(0)}{1+U(0)\Pi \left(
q\right) }\Bigg ],  \label{b5}
\end{eqnarray}
where $q\equiv (\mathbf{Q},\Omega )$ and $\int_{q}\dots \equiv T\sum_{\Omega
_{m}}\int d^{2}q/\left( 2\pi \right) ^{2}\dots $. The last two terms in 
Eq.~(\ref{b5}) correspond to the interaction in the charge and spin channels,
respectively; the first two terms contain unaccounted parts of first- and
second-order diagrams. All terms in Eq.~(\ref{b5}) have perturbative
contributions from the PH continuum. In addition, the charge part contains a
non-perturbative contribution from the ZS pole in the charge-channel
propagator. Near the pole, $\left[ 1-U(0)\Pi (q)\right] ^{-1}\propto
u_{0}^{2}Q^{2}/(\Omega ^{2}-c^{2}Q^{2})$, where $c=v_{F}(1+u_{0}^{2}/2+\dots
)$ is the ZS velocity. Notice that the quanta of ZS are not free bosons: the
residue of the pole vanishes at $Q=0$, as it is required by the
translational invariance of the system.
Substituting the ZS propagator into Eq.~(\ref{b5}), we obtain for the
corresponding contribution to the self-energy
\begin{subequations}
\begin{eqnarray}
\mathrm{Re}\Sigma _{\mathrm{ZS}} &=&\frac{u_{0}^{2}\omega ^{2}}{8E_{F}}%
~F_{R}\left( \frac{\Delta }{\Delta ^{\ast }}\right) ;  \label{zs} \\
\mathrm{I}\text{\textrm{m}}\Sigma _{\mathrm{ZS}} &=&\frac{u_{0}^{2}\omega
^{2}}{4\pi E_{F}}~F_{I}\left( \frac{\Delta }{\Delta ^{\ast }}\right) ,
\label{zs_b}
\end{eqnarray}
where $\Delta ^{\ast }\equiv u_{0}^{2}\omega /2$, and functions $F_{R,I}$
are the real and imaginary parts of
\end{subequations}
\begin{equation}
F(x)=(1+\frac{3}{2}x)\sqrt{1-x}+\frac{3}{2}x^{2}\ln \frac{1+\sqrt{1-x}}{%
\sqrt{-x}},  \label{feb6_7}
\end{equation}
correspondingly. Plots of $F_{R,I}$ are presented in
Fig.~\ref{fig:scaling}. Notice that $F_{I}(x<0)=0$ and
$F_{R}(x>1)=0.$

 For large $x$, \emph{i.e.}, far away from the
mass shell, $F_{I}(x)\propto 1/\sqrt{x}$, and \textrm{Im}$\Sigma _{\mathrm{ZS%
}}\propto u_{0}^{3}\omega ^{5/2}/\sqrt{\Delta }$. This behavior reproduces
the result of the perturbation theory. Expanding further in $1/x$, one obtains
higher powers of $1/\Delta $. However, we see from (\ref{feb6_7}) that the
emerging singularity at $\Delta =0$ is actually cut off at $x\simeq 1$,
\emph{i.e.}, at $\Delta \simeq \Delta ^{\ast }$.
For small $x$, $F_{I}(x)=3\pi
^{2}x^{2}/8$, and  $\text{Im}\Sigma _{\mathrm{ZS}}\propto \Delta ^{2}$ for $%
\Delta \ll \Delta ^{\ast }$. At $\Delta =0$, $\text{Im}\Sigma
_{\mathrm{ZS}}$
 vanishes. Vanishing of Im$\Sigma _{\mathrm{ZS}}$ on the
mass shell is a Cherenkov-type effect:  because the zero-sound velocity $%
c>v_{F}$, an on-shell fermion cannot emit a ZS boson. The $\Delta ^{2}$%
-behavior of $\text{Im}\Sigma _{\mathrm{ZS}}$ in between $\Delta =0$ and $%
\Delta =\Delta ^{\ast }$ tracks an increase of the phase space available for
the emission of ZS bosons. We emphasize that
as $\Delta ^{\ast }\propto u_{0}^{2}$, the
crossover at $\Delta \simeq \Delta ^{\ast }$ could have not been obtained
within the perturbation theory. Near $\Delta=\Delta^*$, the derivative of $\text{Im}\Sigma(\Delta)$
diverges: $\text{Im}\Sigma_I(x)\propto 1/\sqrt{\Delta-\Delta^*}.$  The physical meaning of
scale $\Delta ^{\ast }$ can be understood by noticing that a Cherenkov-type
constraint restricts the frequencies of emitted bosons to the range $%
0<\Omega <\left( \Delta /\Delta ^{\ast }\right) \omega $ (for $\omega >0).$
This constraint is relevant for $\Delta <\Delta ^{\ast }$; for larger $%
\Delta $, emission of bosons with frequencies in the entire interval, allowed
by the Pauli principle ($0<\Omega <\omega )$ is possible.

The real part of the self-energy varies slowly near the mass shell as $%
F_{R}(x\ll 1)=1+x$. Right on the mass shell, Re$\Sigma
_{\mathrm{ZS}}\propto
u_{0}^{2}\omega ^{2}$.
  Notice that $\text{Re}\Sigma _{\text{\textrm{ZS}}}$  is of order $u_{0}^{2}
$, although it is obtained by summing up terms of order $u_{0}^{3}$ and
higher. This enhancement is due to a non-perturbative cut-off of the
mass-shell divergences in the perturbation theory.

The remaining contribution to the forward-scattering part of the self-energy
comes from the PH continuum. It contains a logarithmic divergence to second
order and also power-law divergences to third and higher orders, as within
the perturbation theory the interaction with the continuum is
indistinguishable from the interaction with the zero sound. The logarithmic
divergence in the 
self-energy is cut off by a finite curvature of the spectrum on a scale $%
\Delta \simeq \omega ^{2}/W$, where $W\simeq E_{F}$ is the bandwidth. The
power-law divergences in the PH contribution are cut at the same scale $%
\Delta ^{\ast }$ as in the $\Sigma _{{\rm ZS}}$, this time because
an increase of $U(0){\rm Im}\Pi$ near the boundary of the PH continuum
reduces the effective interaction for $\Delta<\Delta^*$.
However, contrary to the zero-sound contribution,
$\Sigma _{\text{PH}}$ is smooth at $\Delta \simeq \Delta ^{\ast }$
as no Cherenkov-type condition is involved.
Near the mass shell, $\Sigma _{\text{PH}}$ reduces to
\begin{eqnarray}
&&\text{Re}\Sigma _{\text{\textrm{PH}}}=-\frac{u_{0}^{2}|\omega |\Delta }{%
4E_{F}}  \notag \\
\text{Im}\Sigma _{\text{\textrm{PH}}} &=&\frac{u_{0}^{2}\omega ^{2}}{4\pi
E_{F}}\left\{
\begin{array}{ll}
|\ln u_{0}^{2}| & \text{for $|\omega |\ll \omega _{c}$;} \\
|\ln \left( \left| \omega \right| /E_{F}\right) | & \text{for $|\omega |\gg
\omega _{c},$}
\end{array}
\right.   \notag
\end{eqnarray}
where $\omega _{c}\equiv u_{0}^{2}E_{F}$ .

The sum $\Sigma_{{\rm ZS}} + \Sigma_{{\rm PH}}$ is the total contribution to
the self-energy from forward scattering. Another contribution to
the self-energy comes from the processes in which the fermions
move initially in almost opposite directions, and then either
scatter by small angle
(Fig.~\ref{fig:proc}b) or backscatter
(Fig.~\ref{fig:proc}c).
We refer to both  these processes as  ``backscattering''. The
backscattering part of the self-energy, $\Sigma
_{\text{\textrm{B}}},$ is regular on the mass shell and, for weak
interaction, needs to be evaluated only to second order
\cite{chm}. Taking the result for $\Sigma _{\text{\textrm{B}}}$
from Ref.
\cite{chm} and adding it up with the results for $\Sigma _{\text{\textrm{ZS}}%
}$ and $\Sigma _{\text{\textrm{PH}}}$,
we obtain for the total self-energy on the mass shell
\begin{subequations}
\begin{eqnarray}
\mathrm{Re}\Sigma \left( \omega \right) &=&\frac{\omega \left| \omega
\right| }{8E_{F}}\left( u_{0}^{2}-\left[
u_{0}^{2}+u_{2k_{F}}^{2}-u_{0}u_{2k_{F}}\right] \right)  \notag \\
&=&\frac{\omega \left| \omega \right| }{8E_{F}}~u_{2k_{F}}\left(
u_{0}-u_{2k_{F}}\right);  \label{sigma_full_R} \\
\mathrm{Im}\Sigma \left( \omega \right) &=&\frac{u_{0}^{2}}{2\pi }~\frac{%
\omega ^{2}}{E_{F}}\ln \frac{E_{F}}{|\omega |}{\Phi }\left( \frac{\left|
\omega \right| }{u_{\max }^{2}E_{F}}\right) ,  \label{sigma_full_I}
\end{eqnarray}
\end{subequations}
where $u_{2k_{F}}\equiv mU\left( 2k_{F}\right) /2\pi $ and $u_{\max }\equiv
\max \{u_{0},~u_{2k_{F}}\}.$ The scaling function $\Phi \left( x\right) $
approaches the constant values of $%
1(1/2)+u_{2k_{F}}(u_{2k_{F}}-u_{0})/(2u_{0})$, for $x\rightarrow \infty $
and $x\rightarrow 0,$ correspondingly.

The first term in the first line of Eq.~(\ref{sigma_full_R}) is the
non-perturbative contribution from forward scattering, whereas the term in
square brackets is the backscattering contribution. We see that near the
mass shell both contributions are of the same order, \emph{i.e.}, \textit{%
the interaction with the zero-sound collective mode modifies significantly
the perturbative result for the self-energy}. For contact interaction ($%
u_{0}=u_{2k_{F}}$), the zero-sound contribution cancels out with the
backscattering one, so that the net $\mathrm{Re}\Sigma $ vanishes at the
mass shell. At the same time, the effect of the interaction with ZS on the
on-shell $%
\mathrm{Im}\Sigma$  is rather benign: all one has is a
smooth crossover function interpolating between two limiting values of the
prefactor in a familiar $\omega ^{2}\ln \left| \omega \right| $-dependence.
This is the consequence of the fact that the non-perturbative, ZS
contribution to $\mathrm{Im}\Sigma$ vanishes on the mass shell.

\begin{figure}[tbp]
\begin{center}
\epsfxsize=0.7 \columnwidth
\epsffile{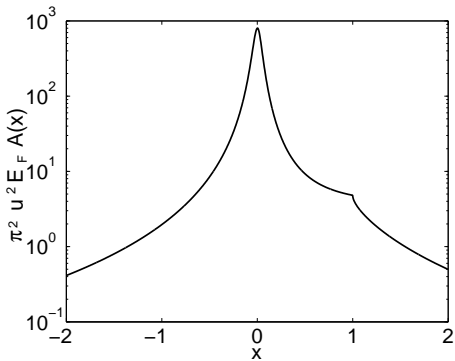}
\end{center}
\caption{A $\log$-plot of the spectral function
$A(\protect\omega,k)$ as a function of
$x=2(\protect\omega-\protect\epsilon_k)/u^2\protect\omega$ for
fixed $\protect\omega$. $L_\protect\protect\omega=2$ and
$\protect\gamma=0.1$ A kink at $x=1$ is a signature of the
interaction with the zero-sound mode.} \label{fig:spectral}
\end{figure}

At the same time, away from the mass shell, the full self-energy
and the the spectral function $A(\omega ,k)=(-1/\pi
)$\textrm{Im}$G\left( \omega ,k\right) $ are affected by the
non-monotonic variation of $\text{Im}\Sigma _{\text{\textrm{ZS}}}$
with $\Delta $ near a threshold for emission of ZS bosons.
 We consider a setup, when $\omega $ is fixed and the spectral
function is measured as a function of $k$.
 This is equivalent to varying $x\equiv \Delta /\Delta ^{\ast }$ at fixed
$\omega $. Near the mass shell, the spectral function consists of two terms $%
A(x)=A_{\text{\textrm{qp}}}(x)+A_{\text{\textrm{ZS}}}(x)$, where
\begin{equation}
A_{\text{\textrm{qp}}}(x)=\frac{1}{\pi ^{2}u^{2}E_{F}}\frac{L_{\omega }}{%
x^{2}+\gamma ^{2}}  \label{A0}
\end{equation}
is a quasi-particle contribution and
\begin{equation}
A_{\text{\textrm{ZS}}}=\frac{F_{I}\left( x\right) }{\pi ^{2}u^{2}E_{F}}\frac{%
x^{2}-\gamma ^{2}}{(x^{2}+\gamma ^{2})^{2}}  \label{AZS}
\end{equation}
is a  ZS contribution.
We
introduced $L_{\omega }\equiv \ln \left( E_{F}/u^{2}|\omega
|\right) $ and $\gamma =|\omega |L_{\omega }/(2\pi E_{F})$, and
assumed for simplicity that $u_{0}\equiv u_{2k_{F}}\equiv u.$
The ZS contribution $A_{\text{\textrm{ZS}}}(x)$ has a sharp maximum at $x=1$%
, where function $F_{I}(x)$ in Fig.~\ref{fig:scaling} has a peak.
This maximum gives rise to a kink in total $A(x)$ (see
Fig.~\ref{fig:spectral}). A similar consideration shows that the
kink is present also for the Coulomb interaction.
This kink should be detectable in
photoemission experiments on layered materials and in
momentum-conserving tunneling between parallel layers of 2D
electron gases \cite{tunneling}.

Finally, we consider an issue whether non-perturbative effects,
discussed above, are relevant for the  specific heat of
two-dimensional fermions. We argue that they are not. The simplest
way to see this is to use the
 definition, $C(T)=-T\partial^2 \Xi/\partial T^2$, and  expand
 the thermodynamic potential $\Xi$ in
powers of $U$. The thermodynamic potential can be
 evaluated in the
Matsubara formalism, in which the polarization operator $\Pi_m =-
(m/2\pi)[1 - |\Omega_m|/\sqrt{(v_F Q)^2+\Omega^2_m}]$ is regular
at any $\Omega_{m}$ and $Q$. As a result, the Matsubara series for
$\Xi$  converges for a weak, short-range interaction, and there is
no need for re-summation of the perturbation theory for $\Xi $.
Evaluating $\Xi$ to second-order in $U$ yields
\begin{equation}
\delta C\left( T\right) =-\left(
u_{0}^{2}+u_{2k_{F}}^{2}-u_{0}u_{2k_{F}}\right) ~\frac{3 m\zeta (3)}{\pi }~%
\frac{T^2}{E_{F}}.  \label{genu}
\end{equation}
This result coincides with $C(T)$
 obtained in Ref.~\cite{chm} by finding the perturbative, backscattering
part of the self-energy first and then using the relation between
$C(T)$ and $\Sigma$ \cite{agd}. The agreement between the two results shows
 that non-parturbative, forward-scattering self-energy does not affect $C(T)$.

For completeness, we also verified explicitly that there is no
  contribution to $C(T)$ from forward scattering.
 To this end, one can  use the relation between $C(T)$
  and the Green's function \cite{agd} which, to first order in
  $(\omega-\epsilon_k)^{-1}\Sigma(\omega,k)$, reads
\begin{eqnarray}
&&\delta C(T)=\frac {mT}{\pi}\frac{\partial}{\partial T}\Biggl[\frac{1}{T}%
\int^{\infty}_{-\infty} d\epsilon_k\int_{-\infty}^{\infty} d\omega
\omega
\frac{\partial n_0}{\partial \omega}  \label{c2} \\
&&\times\Bigg\{ \delta(\omega-\epsilon_k)\text{Re}\Sigma^R(\omega,k) - \frac{%
1}{\pi}\mathcal{P}\frac{1}{\omega-\epsilon_k} \text{Im}\Sigma^R(\omega,k)%
\Bigg\} \Biggr].  \notag
\end{eqnarray}
Here $n_0(\omega)$ is the Fermi function. It is crucial that both
$\text{Re}\Sigma^R(\omega,k)$ and $\text{Im}\Sigma^R(\omega,k)$
are present in Eq.~(\ref{c2}). The forward scattering self
energy, $\Sigma_{\rm F} = \Sigma_{{\rm ZS}} + \Sigma_{{\rm PH}}$, depends on $k$ due to the presence
of a non-perturbative scale $\Delta^*$, and therefore
$\text{Im}\Sigma_{\rm F}$ yields a finite contribution to
$C(T)$. Substituting the results for $\Sigma_{{\rm ZS}}$ and $\Sigma_{{\rm PH}}$
 into Eq.~(\ref{c2}), we find
that the forward scattering contributions to the specific heat from
$\text{Re}\Sigma_{{\rm F}}$ and $\text
{Im}\Sigma_{{\rm F}}$ cancel each other; hence, there is no contribution to $C(T)$
 from forward scattering.

We acknowledge stimulating discussions with I. Aleiner, B. Altshuler, A.
Andreev, W. Metzner, A. Millis, and C. Pepin. The research has been supported by NSF
DMR 0240238 and Condensed Matter Theory Center at UMD
 (A. V. Ch.), NSF DMR-0308377 (D. L. M.), and
NSF DMR-0237296 (L. I. G.).


\vspace{-0.3cm}

\begin{thebibliography}{*}
\bibitem[*]{perm}  Permanent address.

\bibitem{agd}  A.\ A.\ Abrikosov, L.\ P.\ Gorkov, and I.\ E.\ Dzyaloshinski,
\emph{Methods of quantum field theory in statistical physics}, (Dover
Publications, New York, 1963).

\bibitem{anderson}  P. W. Anderson, Phys. Rev. Lett. \textbf{65, }2306
(1990).

\bibitem{castellani}  C. Castellani, C. Di Castro, and W. Metzner, \prl {72}%
, 316 (1994).

\bibitem{fukuyama}  H. Fukuyama and M. Ogata, J. Phys. Soc. Jpn. \textbf{63}%
, 3923 (1995).

\bibitem{metzner}  C. Halboth and W. Metzner, \prb {\bf 57}, 8873 (1998).

\bibitem{chm}  a) A. V. Chubukov and D. L. Maslov, Phys. Rev. B \textbf{68, }%
155113 (2003); b) \emph{ibid. }\textbf{69}, 121102 (2004).

\bibitem{1D}  Yu. A. Bychkov, L. P. Gor'kov, and I. E. Dzyaloshisnkii, Zh.\
Eksp. Teor. Fiz. \textbf{50}, 738 (1966) [Sov.\ Phys.\ JETP \textbf{23},
489].

\bibitem{bedell}  D. Coffey and K. S. Bedell, Phys. Rev. Lett. \textbf{71},
1043 (1993).

\bibitem{dassarma}  V. M. Galitski and S. Das Sarma, \prb {\textbf 70}, 035111 (2004).
\bibitem{long}  A. V. Chubukov, D. L. Maslov, S. Gangadharaiah, and L. I.
Glazman, cond-mat/0412283.

\bibitem{comm}  In principle, there is no guarantee that Eq.~(\ref{c2})
works for non-linear terms in $C(T)$. However, it can be verified by a direct
comparison with the general Luttinger-Ward formula that the $T^2$-term is
correctly described by (\ref{c2}) if one uses the $T$-dependent self-energy rather than the
zero-temperature one.


\bibitem{tunneling}  see S. Q. Murphy, J. P. Eisenstein, L. N. Pfeiffer, and
K. W. West, Phys. Rev. B \textbf{52}, 14825-14828 (1995) and references
therein.
\end{thebibliography}
\end{document}